\documentclass[useAMS,usenatbib]{mn2e}

\usepackage{savesym}
\usepackage{graphicx}
\usepackage{fixltx2e} 
\usepackage{dcolumn}
\usepackage[version=3]{mhchem} 
\usepackage{natbib}
\usepackage{color}
\usepackage{latexsym}

\newcommand{\core}{G333.125--0.562}
\newcommand{\cohigh}{CO (4 -- 3)}
\newcommand{\colow}{CO (1 -- 0)}
\newcommand{\kms}{km s$^{-1}$}
\newcommand{\um}{${\rm \mu m}$}
\newcommand{\solarM}{M$_\odot$}

\title[Modelling of massive dense cold core G333.125--0.562]{Observations and radiative transfer modelling of a massive dense cold core in G333}

\author[Lo et al.]{N.~Lo$^{1,2,3,4}$\thanks{current address: Universidad de Chile, nlo@das.uchile.cl}, M.P.~Redman$^{5}$, P.A.~Jones$^{1,2}$, M.R.~Cunningham$^{2}$, R.~Chhetri$^{2,4}$, \and I.~Bains$^{6}$ and M.G.~Burton$^{2}$\\
$^{1}$Departamento de Astronom\'ia, Universidad de Chile, Camino El Observatorio 1515, Las Condes, Santiago, Casilla 36-D, Chile\\
$^{2}$School of Physics, University of New South Wales, Sydney, NSW 2052, Australia\\
$^{3}$Laboratoire AIM Paris-Saclay, CEA/Irfu - Uni. Paris Did\'erot - CNRS/INSU, 91191 Gif-sur-Yvette, France\\
$^{4}$Australia Telescope National Facility, CSIRO Astronomy and Space Science, Sydney, NSW 2052, Australia\\
$^{5}$Centre for Astronomy, School of Physics, National University of Ireland, Galway, University Road, Galway, Ireland\\
$^{6}$Centre for Astrophysics and Supercomputing, Swinburne University of Technology, P.O. Box 218, Hawthorn, VIC 3122, Australia\\}

\begin{document}

\date{Accepted ***. Received ***; in original form ***}
\pagerange{\pageref{firstpage}--\pageref{lastpage}} \pubyear{}
\maketitle
\label{firstpage}
 
\begin{abstract}
Cold massive cores are one of the earliest manifestations of high mass star formation. Following the detection of SiO emission from G333.125--0.562, a cold massive core, further investigations of the physics, chemistry and dynamics of this object has been carried out. Mopra and NANTEN2 molecular line profile observations, Australia Telescope Compact Array (ATCA) line and continuum emission maps, and Spitzer 24 and 70 \um\ images were obtained. These new data further constrain the properties of this prime example of the very early stages of high mass star formation. A model for the source was constructed and compared directly with the molecular line data using a 3D molecular line transfer code - {\sc MOLLIE}. The ATCA data reveal that G333.125--0.562 is composed of two sources. One of the sources is responsible for the previously detected molecular outflow and is detected in the Spitzer 24 and 70 \um\ band data. Turbulent velocity widths are lower than other more active regions of G333 which reflects the younger evolutionary stage and/or lower mass of this core. The molecular line modelling requires abundances of the CO isotopes that strongly imply heavy depletion due to freeze-out of this species onto dust grains. The principal cloud is cold, moderately turbulent and possesses an outflow which indicates the presence of a central driving source. The secondary source could be an even less evolved object as no apparent associations with continuum emissions at (far-)infrared wavelengths.
\end{abstract}


\section{Introduction}
High mass star formation is unlikely to be simply a scaled up version of low mass star formation because the final outcomes are dramatically different. Low mass star formation produces isolated or loosely clustered stars while high mass star formation yields dense clusters of thousands of low mass stars alongside relatively few massive stars in an ionized nebula. Understanding how high mass star formation is initiated and proceeds requires observations of the earliest stages of the process, before the newly formed massive star has completely disrupted its surroundings. Objects that appear to be high-mass analogues of the Class 0 \citep*{Andre1993} sources observed in low mass star formation are now being uncovered \citep{Wu2005, Garay2004, Birkmann2007}. These objects can be studied to determine how close an analogue of Class 0 sources they are, enabling the origins of the differences between low and high mass star formation.

\core\ was originally noted as a potentially young object by \citet{Garay2004} and measurements of the mass and temperature of the source by \citet{Lo2007} confirmed the source to be massive dense and cold. It has a gas mass of $1.8 \times 10^3$ \solarM, temperature of 13 to 19 K and is undetected at wavelengths shorter than 24 \um. \citet{Lo2007} found \core\ to be a source of SiO emission, and the broader width of the SiO line compared to other species indicates that this shocked gas tracer species is likely to have been excited in a molecular outflow. This finding is interesting in that it suggests that the central source must be evolved enough for an organised infall and accretion via an envelope(and/or a disk) to have already developed, yet before the source is readily detectable in the infrared.

To further investigate \core\ we have obtained new single dish (CO J = 4 -- 3) and interferometric molecular line (\ce{HCO+} $J = 1 - 0$, \ce{SiO} $J = 1 - 0$ and $J = 2 - 1$) data, which are presented in \S \ref{sec:observations} and \S \ref{sec:results}. The integrated emission map combined with Spitzer MIPS data reveal that \core\ actually contains two components. One of these sources dominates the emission and drives the outflow while the nature of the other source is less clear, being either an extension of \core\ or perhaps an even younger cold core. Maps of molecular line profiles from spatial positions across the source are presented and these data are modelled using a 3D molecular line radiative transfer code - {\sc MOLLIE} (MOLecular LIne Explorer; \S \ref{sec:modelling}). The modelling indicates that heavy depletion of CO and its isotopes is taking place, and the abundances required for all species are consistent with the results of \citet{Lo2007}. In \S \ref{sec:discussion} the source is compared and contrasted with other sites of high mass star formation within the G333 complex. Conclusions are drawn in \S \ref{sec:conclusions}.

\section{Observations} \label{sec:observations}
Data presented in this work were collected throughout 2004 to 2008, comprised of data from the Mopra Radio Telescope, the Australia Telescope Compact Array (ATCA)\footnote{http://www.atnf.csiro.au} and the NANTEN2 Observatory\footnote{http://www.astro.uni-koeln.de/nanten2}. Observed molecular transitions and continuum emission are listed in Table \ref{tab:line_list}. 

\begin{table*}
  \caption{List of observed molecular transitions and continuum of \core\ presented in this paper. The columns are: (1) molecule; (2) transition; (3) rest frequency \citep*{Lovas1979}; (4) upper energy levels from the Cologne Database for Molecular Spectroscopy \citep[CDMS,][]{{Muller2001},{Muller2005}}; (5) calculated critical density: $n_{crit}=A_{ul}/\langle\sigma(v)v\rangle$ where $A_{ul}$ is the Einstein A coefficient obtained from CDMS, $\langle\sigma(v)\rangle=10^{-15}$ cm$^2$ is the collision cross section and $v$ is the velocity of collision particles assumed to be 1 km s$^{-1}$; (6) observation season; (7) telescope used and (8) the rms noise level.}
  \label{tab:line_list}
  \newcolumntype{.}{D{.}{.}{-1}}
  \begin{minipage}{1\textwidth}
  \centering
  \begin{tabular}{c c . c c c c c c}
  \hline
  Molecule & Transition & \multicolumn{1}{c}{Rest frequency} & $E_u/k$ & $n_{crit}$
   & Observation season & Telescope & rms \\
   & & \multicolumn{1}{c}{\scriptsize (MHz)} & \scriptsize (K) & \scriptsize (cm$^{-3}$) & & & \\
  \hline \hline
  \ce{^{13}CO} & $1-0$ & 110201.353 & 5.29 & $3\times10^3$ & 2004 & Mopra & 0.3 K ($T^*_A$) channel$^{-1}$ \\
  \ce{C^{18}O} & $1-0$ & 109782.173 & 5.27 & $3\times10^3$ & Jul, 2005 & Mopra & 0.1 K ($T^*_A$) channel$^{-1}$ \\
  CO & $4-3$ & 461040.768 & 55.31 & $6\times10^6$ & Dec, 2008 & NANTEN2 & 0.5 K ($T^*_A$) channel$^{-1}$ \\
  CS & $2-1$ & 97980.953 & 7.05 & $2\times10^5$ & Sep, 2006 & Mopra & 0.1 K ($T^*_A$) channel$^{-1}$ \\
  \ce{C^{34}S} & $2-1$ & 96412.961 & 6.94 & $2\times10^5$ & Sep, 2006 & Mopra & 0.1 K ($T^*_A$) channel$^{-1}$ \\
  \ce{HCO+} & $1-0$ & 89188.526 & 4.28 & $4\times10^5$ & Jul, 2006 & Mopra & 0.1 K ($T^*_A$) channel$^{-1}$ \\
   & & & & & Oct, 2007 & ATCA & 0.1 K channel$^{-1}$ \\
  \ce{H^{13}CO+} & $1-0$ & 86754.330 & 4.16 & $4\times10^5$ & Jul, 2006 & Mopra & 0.1 K ($T^*_A$) channel$^{-1}$\\
  SiO & $1-0\,\nu=0$ & 43423.853 & 2.08 & $3\times10^5$ & Mar, 2008 & ATCA & 0.2 K channel$^{-1}$ \\
  SiO & $2-1\,\nu=0$ & 86847.010 & 6.25 & $3\times10^5$ & Jul, 2006 & Mopra & 0.1 K ($T^*_A$) channel$^{-1}$\\
   & & & & & Oct, 2007 & ATCA \\
  \multicolumn{2}{c}{7-mm continuum} & \multicolumn{1}{c}{43440\footnote{Centre frequency of the observation window.}} & & & Mar, 2008 & ATCA & 0.004 Jy beam$^{-1}$ \\
  \multicolumn{2}{c}{3-mm continuum} & \multicolumn{1}{c}{86858$^a$} & & & Oct, 2007 & ATCA & 0.05 Jy beam$^{-1}$ \\
  \hline
  \end{tabular}
  \end{minipage}
\end{table*}

The mapping data were collected between 2004 July and 2006 October with the Mopra Telescope, as part of the multi-molecular line mapping of the G333 complex \citep[for details see][]{Bains2006, Wong2008, Lo2009}. The Mopra Telescope is a 22-metre telescope located near Coonabarabran, Australia. It is a centimetre- and millimetre-wavelength antenna having a full width to half-maximum (FWHM) beam size of $\sim$36 arcseconds and velocity resolution of $\sim$0.1 \kms\ channel$^{-1}$ at 100-GHz \citep{Ladd2005}. The observations were carried out with the University of New South Wales Mopra Spectrometer (UNSW-MOPS) digital filterbank backend and Monolithic Microwave Integrated Circuit (MMIC) receiver, except for \ce{^{13}CO} and \ce{C^{18}O} which were observed with the Mopra Correlator (MPCOR) and the previous SIS receiver. For full observational setup details, see \citet{Bains2006, Wong2008, Lo2009}.

High resolution SiO, \ce{HCO+} and continuum (3 and 7 mm) imaging of the core were obtained with the ATCA, which is an array of six 22-metre antennas at the Paul Wild Observatory, Australia. The data was collected during 2007 October and 2008 March, with array configuration of H75 and H168, giving a synthesis beam size of $6\arcsec \times 4\arcsec$ and $5\arcsec \times 3\arcsec$ at 3 and 7 mm wavelengths respectively. For the 3-mm observations only five antennas were in use. The primary flux calibrator used was Uranus and the secondary calibrator was 1646-50 for both 3 and 7 mm observations. The flux of Uranus were 7.5 and 1.9 Jy at 3 and 7 mm wavelengths respectively. The flux of 1646-50 were 1.6 and 1.9 Jy at 3 and 7 mm wavelengths respectively. The spectral resolution is 0.8 \kms\ per channel for both wavelengths. ATCA data were reduced with the radio interferometry data reduction package {\sc miriad}\footnote{http://www.atnf.csiro.au/computing/software/miriad}. All ATCA images presented in this work are corrected for the primary beam area, the Jansky to Kelvin conversion factors are 7.2 and 46.5 K Jy$^{-1}$ for 3 and 7 mm data respectively. 

CO (4 -- 3) spectra were obtained with the NANTEN2 Telescope at Pampa la Bola, Chile. The telescope is 4 metres in diameter, with angular resolution of 45 arcseconds, velocity resolution of 0.6 \kms\ channel$^{-1}$ and beam efficiency of 0.5 at 460-GHz. Observations were made with the KOSMA SMART receiver, a dual-frequency $2 \times 8$ pixel array receiver and the KOSMA array Acousto Optical Spectrometers (AOS) as backends.

The 24 and 70 \um\ images presented in this work are part of the Spitzer legacy project `A 24 and 70 Micron Survey of the Inner Galactic Disk with MIPS' \citep[MIPSGAL; ][]{Carey2009}. The resolutions are 6 and 18 arcseconds at 24 and 70 \um\ respectively. The 5$\sigma$ rms point source sensitivities are 1.3 and 73 mJy at 24 and 70 \um.\\

\section{Results} \label{sec:results}
\subsection{Continuum} \label{sec:results_cont}
\begin{figure}
  \centering
  \includegraphics{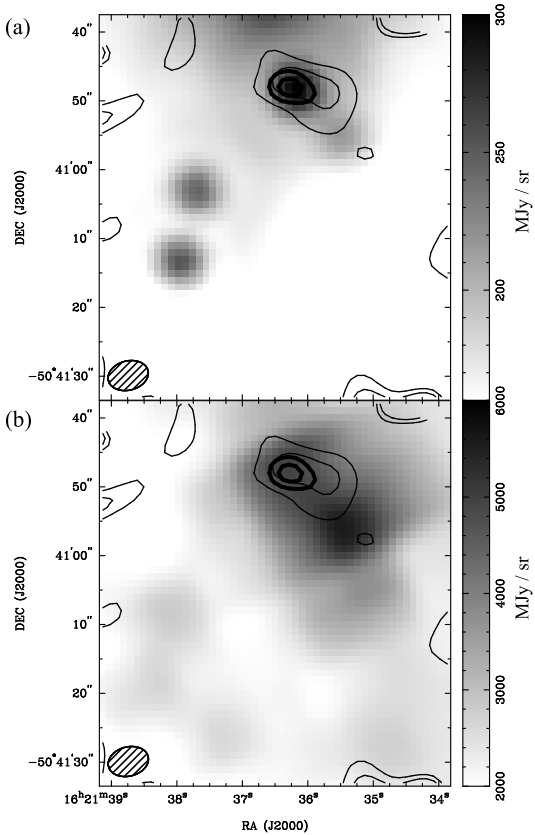}
  \caption{(a) ATCA 3 and 7 mm continuum (thin and thick contours respectively) overlain on Spitzer MIPS 24-\um\ continuum image of \core. The contour levels are at 0.03 and 0.06 Jy beam$^{-1}$ for the 3-mm continuum, 0.002 and 0.003 Jy beam$^{-1}$ for the 7-mm continuum. (b) Same contoured data but overlaid on MIPS 70-\um\ image. The crossed ellipse at lower left corner represent the beam size of the 3 and 7 continuum data.}
  \label{fig:continuum}
\end{figure}

\begin{figure}
  \centering
  \includegraphics{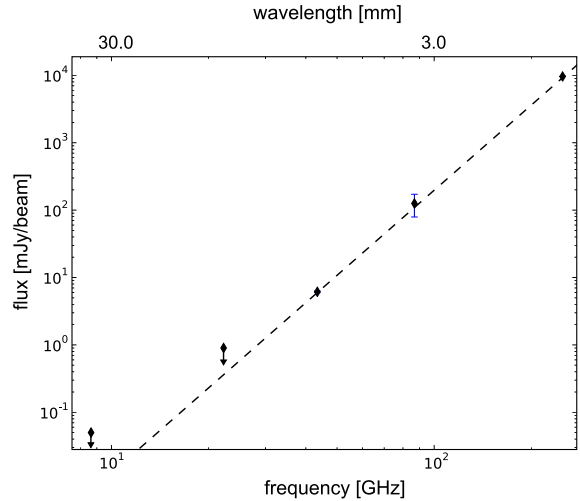}
  \caption{A plot of continuum peak flux-frequency of the core. The 3-cm \citep{Jones2008a} and 12-mm continuum data points are the upper limit, the 1.2-mm data is from \citet{Mookerjea2004}. The plot shows a spectral index $\alpha$ of 4.2 for the three millimetre data points (dashed line) and no clear sign of turning point.}
  \label{fig:spectral_index}
\end{figure}

Figure \ref{fig:continuum} is a map of the ATCA interferometric image of 3 and 7 mm continuum (thin and thick contours respectively), overlain on the MIPSGAL 24-\um\ (a) and 70-\um\ (b) image of \core. The millimetre continuum emission coincides with the 24 and 70 \um\ emission, while the 70-\um\ image reveals that there is another stronger source at approximately 5 arcseconds offset to the  southwest, with fainter 24-\um\ emission. Hereafter we refer to the stronger 24-\um\ source as the `core'. From \citet{Jones2008a}, the core is not detectable at 21, 13, 6 and 3 cm wavelengths. Figure \ref{fig:spectral_index} is a plot of continuum flux-frequency of the core with the 12-mm and 3-cm data points being the upper limit. The spectral index $\alpha$ ($S_\nu \propto \nu^\alpha$, the slope of flux-frequency plot) calculated using the integrated fluxes at 1.2 \citep{Mookerjea2004}, 3 and 7 mm is $4.20 \pm 0.07$. Since these points lie on the Rayleigh-Jeans approximation of the Planck function, the index should give a good approximation of the actual spectral index. The derived spectral index $\alpha$ yields a dust opacity spectral slope of $\beta = \alpha - 2 = 2.20 \pm 0.07$. There is no clear indication of a turnover frequency (where the opacity changes from thick to thin) from the flux-frequency plot. The implication of this is discussed in \S \ref{sec:discussion}. 

\subsection{Molecular lines} \label{sec:result_mol}
\begin{figure}
  \centering
  \includegraphics{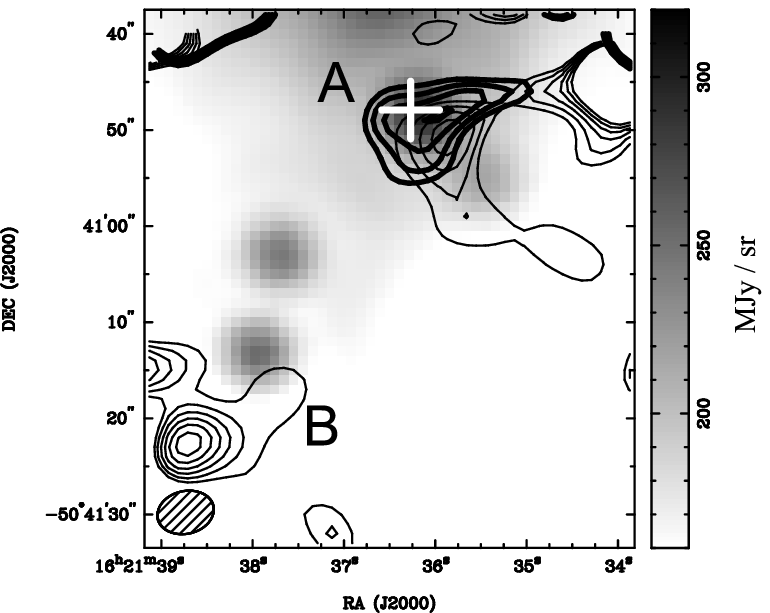}
  \caption{Integrated intensity map of \ce{HCO+} (1 -- 0; thin contours) and SiO (2 -- 1; thick contours) overlain on Spitzer MIPS 24-\um\ image (grey scale) of \core. Integrated velocity range is from $-80$  to $-50$ \kms\ for both \ce{HCO+} and \ce{SiO}. The lowest contour level is 25 K \kms\ with increments of 5 K \kms\ for \ce{HCO+} and 20 K \kms\ with increments of 5 K \kms\ for SiO. The cross marks the position of the 3 and 7 mm continuum peaks, and the crossed ellipse represents the beam size of the molecular line data. The two distinctive molecular emission peaks are denoted by `A' and `B'.}
  \label{fig:intensity}
\end{figure}

\begin{figure}
  \centering
  \includegraphics{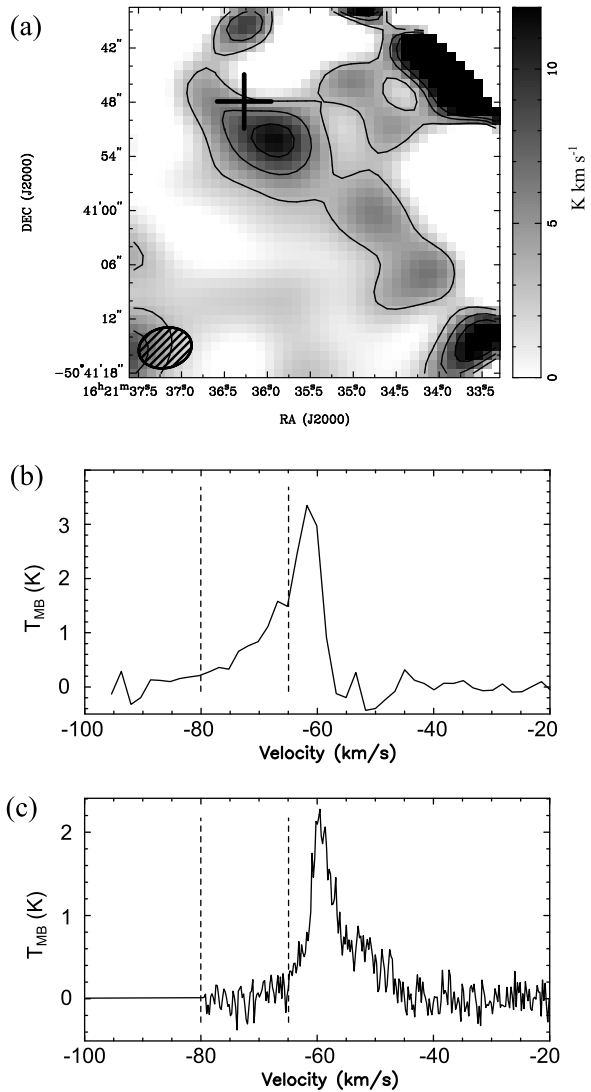}
  \caption{(a) Integrated emission map of the ATCA measurement of the \ce{HCO+} line towards source `A', over a velocity range of $-64$ to $-70$ \kms, showing the blue-shifted line wing contributing to the outflows that is not seen in the Mopra data. The contour levels are at 3.3, 6.6 and 9.9 K beam$^{-1}$. (b) Line profile at the peak \ce{HCO+} (ATCA) position showing the blue-shifted line wing. (c) \ce{HCO+} spectra obtained with Mopra showing the absence of blue-shifted line wing. The dashed lines highlight the blue-shifted line wings velocity range.}
  \label{fig:atca_HCOp_blue}
\end{figure}

Shown in Figure \ref{fig:intensity} is an ATCA map of the zeroth moment (integrated intensity) of the \ce{HCO+} (1 -- 0) data (thin contours) and \ce{SiO} (1 -- 0) data (thick contours), overlain on a MIPS 24-\um\ image (grey scale) of the core. The elongated shape of the \ce{HCO+} emission comes from the blue-shifted line wings of molecular outflow as shown in Figure \ref{fig:atca_HCOp_blue}a, an integrated emission map over a velocity range of $-64$ to $-70$ \kms. The ATCA \ce{HCO+} spectra at the peak position (Figure \ref{fig:atca_HCOp_blue}b) clearly shows the blue-shifted line wing not seen in the Mopra \ce{HCO+} spectra (Figure \ref{fig:atca_HCOp_blue}c). No red component of the outflow is seen in the ATCA data, which is likely spatially oriented such that it is outside the primary beam. Further details are presented in Section \ref{sec:modelling} on the modelling of \ce{HCO+} lines. 

Several 24-\um\ sources are clearly present yet the Spitzer 70-\um\ emission (Figure \ref{fig:continuum}) is confined to the brighter source to the northwest, marked `A' in Figure \ref{fig:intensity}. The southeast \ce{HCO+} emission is marked `B'. Furthermore, the SiO (2 -- 1) integrated intensity data (thick contours) is also confined to the northwest source. Based on its association with the SiO emission detected by \citet{Lo2007} and the spectral energy distribution, the northwest source (`A') was identified as the primary source in this study. The nature of the secondary source and its possible association with \core\ is discussed in Section \ref{sec:discussion}. 

\begin{figure*}
  \centering
  \includegraphics{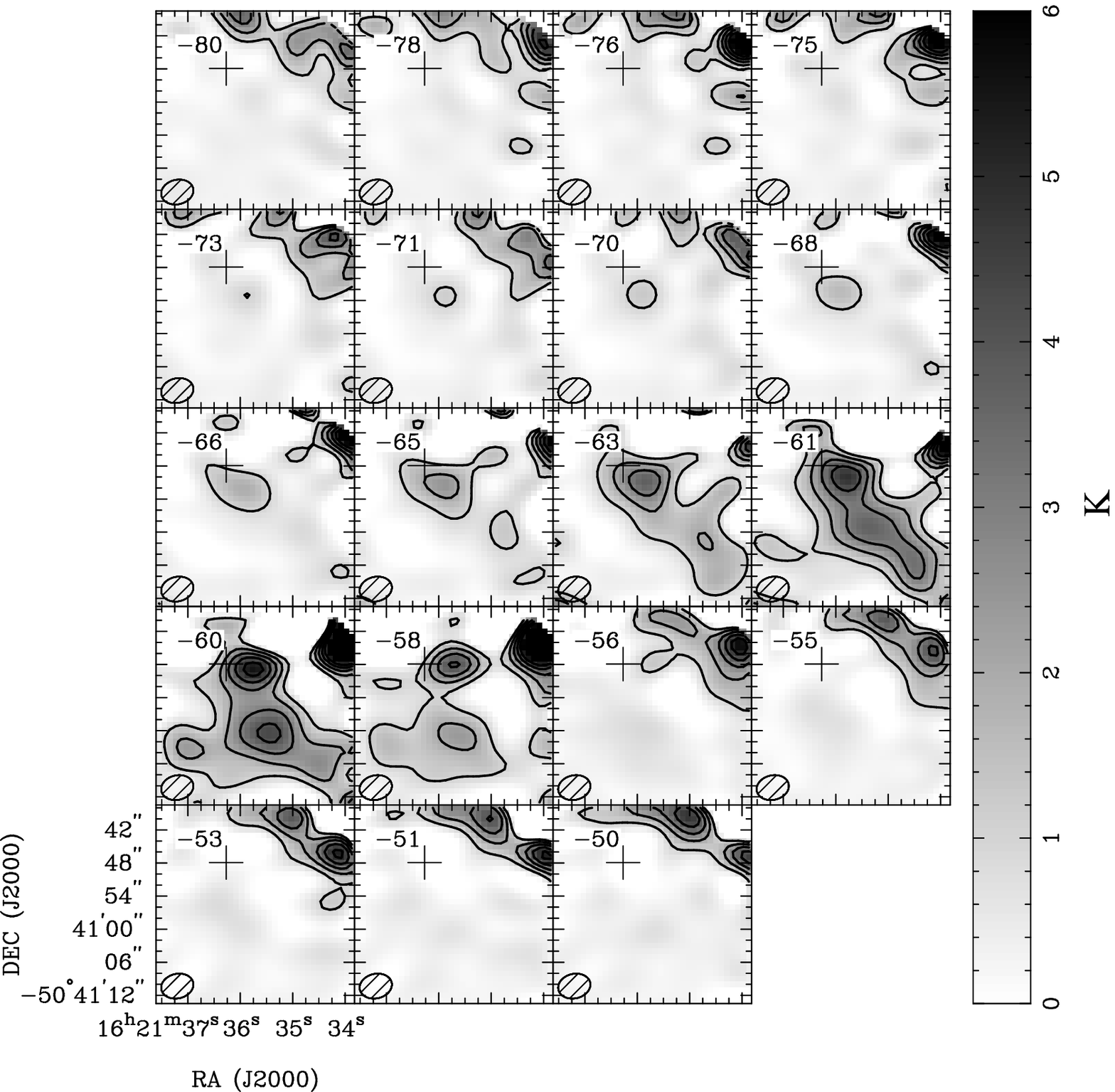}
  \caption{Channel maps of ATCA \ce{HCO+} (1 -- 0) of the core. The lowest contour level is at 1.0 K with increments of 1.0 K. The cross marks the position of 3 and 7 mm continuum, and the crossed ellipse represents the beam size of the molecular line data.}
  \label{fig:hcop_chanmap}
\end{figure*}

\begin{figure*}
  \centering
  \includegraphics{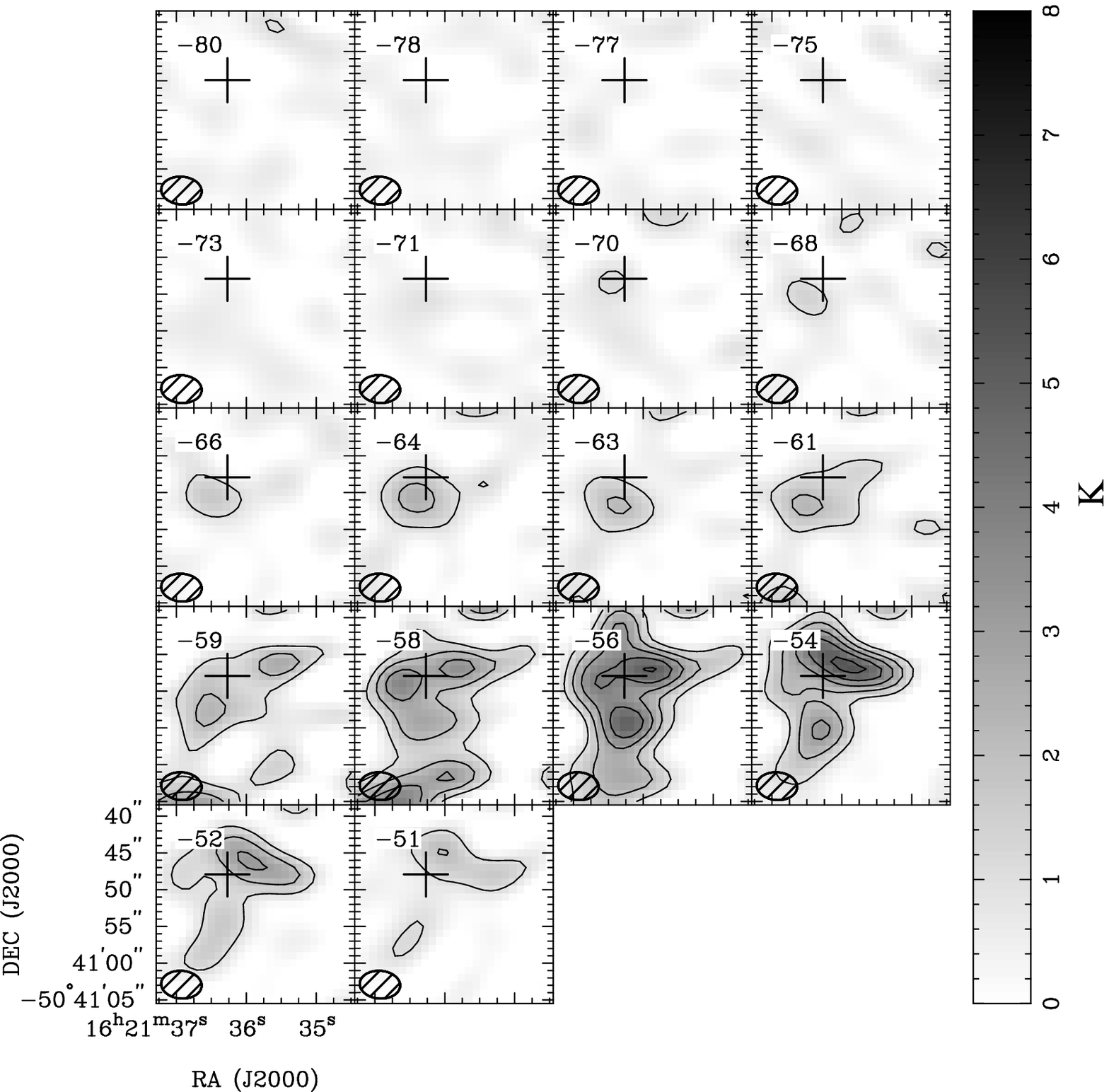}
  \caption{Channel maps of ATCA \ce{SiO} (1 -- 0) of the core. Contour level starts at 1.0 K with increments of 1.0 K. The cross marks the position of 3 and 7 mm continuum, and the crossed ellipse represents the beam size of the molecular line data.}
  \label{fig:sio_chanmap_1}
\end{figure*}

\begin{figure*}
  \centering
  \includegraphics{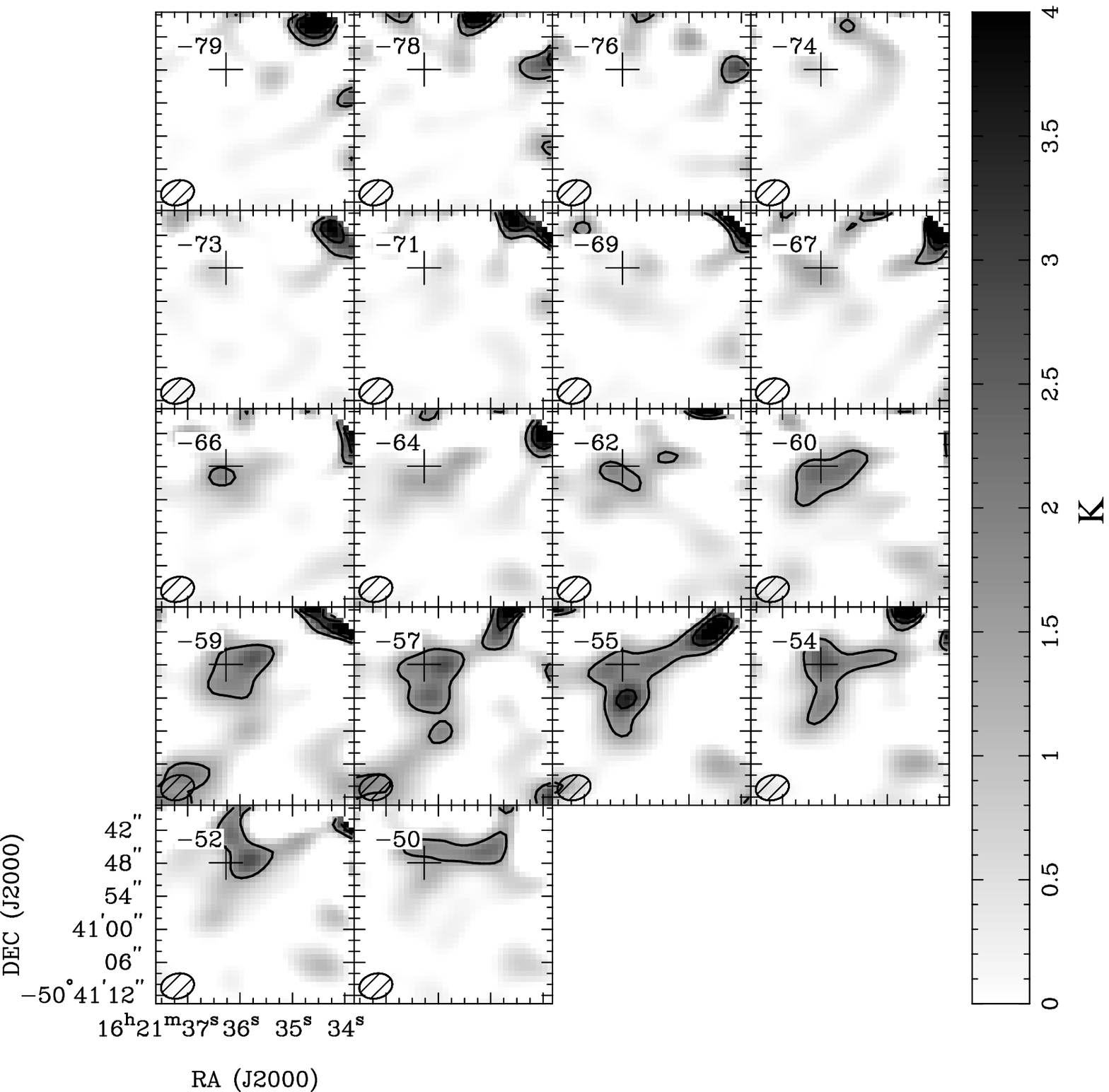}
  \caption{Channel maps of ATCA \ce{SiO} (2 -- 1) of the core. Contour level starts at 1.5 K with increments of 1.5 K. The cross marks the position of 3 and 7 mm continuum, and the crossed ellipse represents the beam size of the molecular line data.}
  \label{fig:sio_chanmap_2}
\end{figure*}

Note that the \ce{HCO+} spectrum does not show clear signs of infall (asymmetry profile) unlike the lower resolution Mopra data; the implication of this is discussed in Section \ref{sec:discussion}. Figure~\ref{fig:hcop_chanmap}, \ref{fig:sio_chanmap_1} and \ref{fig:sio_chanmap_2} show the channel maps of the ATCA \ce{HCO+} and SiO data respectively. 

\begin{figure}
  \centering
  \includegraphics{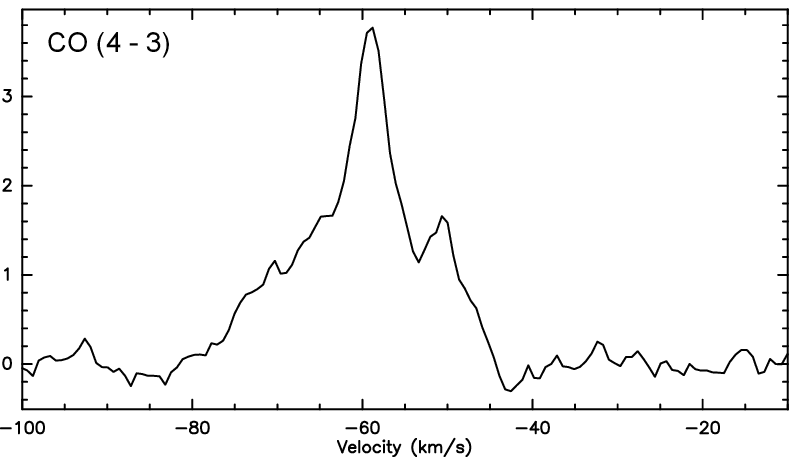}
  \caption{\cohigh\ line profile of the core obtained with NANTEN2. The line profile shows two emission peaks at $-59$ and $-50$ \kms\ .}
  \label{fig:cohigh_spec}
\end{figure}

Figure~\ref{fig:cohigh_spec} show the line profile of \cohigh\ obtained with the NANTEN2 telescope,  with a beam size of 45 arcseconds. The line profile shows two emission peaks at $-59$ and $-50$ \kms\ where the G333 cloud itself contributes to the $-50$ \kms\ peak. Blue-shifted emission is also present, from $\sim -65$ to $\sim -80$ \kms, consistent with the blue shifted emission in the ATCA spectrum (Figure \ref{fig:atca_HCOp_blue}), while the red-shifted component is confused with the $-50$ \kms\ ambient cloud emission. \cohigh\ is a moderate excitation line that should better trace the warmer central regions of \core\ than lower excitation lines such as \colow\, which typically contain contributions from low density emission along the line of sight through the molecular cloud complex.

Finally, using new and existing Mopra and ATCA data, spatial grids of \ce{HCO+} line profiles from positions across the face of \core\ were produced. These line profiles encode information about the velocity, density, turbulence and temperature fields inside the core. In the next section, a fully 3D molecular line profile radiative transfer code is used to closely model this set of line profiles and so the data is presented alongside the model line profiles.

\section{Radiative transfer modelling} \label{sec:modelling}
To verify and extend the observational analysis, a 3D molecular line transport radiative transfer code - {\sc MOLLIE} was used to model the primary source in \core. {\sc MOLLIE} takes in a 3D grid of physical parameters (e.g. density, temperature, velocity, turbulence, chemical abundance) of a source, then simulates the molecular emission profiles for comparison with observed line profiles. Examples of the use of the code can be found in \citet{Keto2004,Redman2006,Carolan2008,Carolan2009}. A model of the source was constructed based on the observational parameters (chemical abundance, density, velocity, gas temperature and line intensity) from \citet{Lo2007} and references therein, with the aim of reproducing line profiles of \ce{^{13}CO} (1 -- 0), \ce{C^{18}O} (1 -- 0), CS (2 -- 1) and \ce{HCO+} (1 -- 0), and using the NANTEN2 CO (4 -- 3) transition as a constraint to the model. 

We have generated over a thousand models comprised of different physical parameters that give rise to prominent features in the line profiles: density and abundance (line strength), temperature (line strength and line widths), infall/expansion velocity (asymmetric self-absorption line), outflow velocity (line wings). By fitting multiple molecular transition line profiles, the number of possible best fit models is greatly constrained, leaving only a handful to choose from. The physical parameters of these limited number of possible models vary within a small threshold (a factor of two). The modelling consists of two parts. The first reduces the number of possible models by fitting the line profiles of multiple Mopra molecular lines. In the second part we refine the model with high resolution ATCA data. Details of the modelling is described in the following paragraphs.

The radius of the core was taken to be $R = 0.5~{\rm pc}$. The temperature increases linearly from the edge of the core (12 K) to the centre (17~K). The density distribution is modelled as a simple Plummer sphere $\rho=(\rho_{\rm c} R_{\rm p}^2)/(R_{\rm p}^2+r^2)$ such that the density is approximately constant in the centre of the core at $\rho_{\rm c}=5 \times 10^5~{\rm cm^{-3}}$ and then falls off as $r^{-2}$. The scaling factor of $R_{\rm p}=0.1R$ is quite small which means that the density distribution is close to a $r^{-2}$ power law. This density distribution was chosen to be consistent with the fall off in the strength of the line profiles away from the centre of the core, as well as with the dust emission profile.

\subsection{Modelling of Mopra data} \label{sec:modelling_mopra}
\begin{figure}
  \centering
  \includegraphics{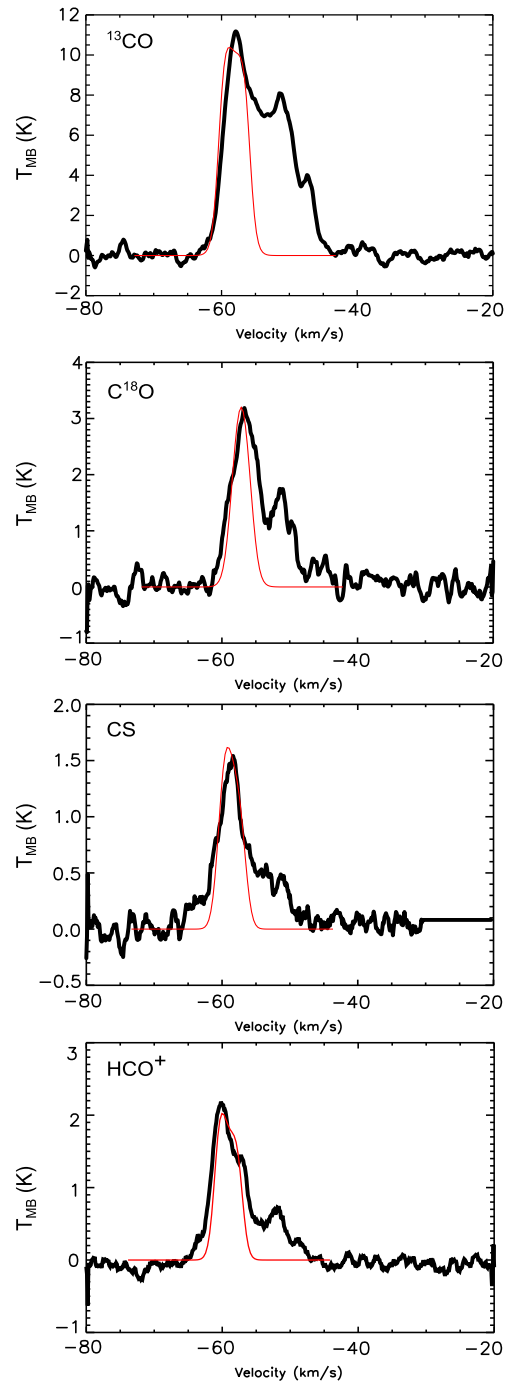}
  \caption{Modelled line profiles (thin lines) and Mopra observations line profiles (thick lines) of the core. We aim to reproduce the line strength and line width of the observed molecular transitions. Note the observed emissions at $\sim -50$ \kms\ is due to ambient cloud emissions and thus not present in the model spectra.}
  \label{fig:cc_mopra_model}
\end{figure}

Typical line widths for species other than SiO are 4 - 5 \kms. Thermal broadening (due to gas at 12 - 17 K) is far too small to generate these line widths which implies that dynamical processes such as turbulence, infall, outflow and rotation must be responsible for line widths. Accordingly, the model also incorporates a directed molecular outflow, radial infall and bulk turbulence. Infall is required because lines such as \ce{HCO+} (Mopra data) exhibit a self-absorption profile with a stronger blue-shifted peak to that observed in nearby low mass infall sources. The line split due to infall is marginally seen in the asymmetrical shape of the line profile, in which the blue shifted emission is stronger due to self-absorption of the red emission (Figure \ref{fig:cc_mopra_model}), suggesting the infall velocity is comparable to the turbulent velocity. The self-absorption of the red shifted emission in this way, rather than widely split or blended profiles, is routinely seen in single dish observations of infall in low mass star forming clouds, and is seen here with the larger line widths and splitting of a massive star forming region. Isotopes such as \ce{H^{13}CO+} show single peaked Gaussian shape profiles and turbulence is required to give a structureless broad line profile at these temperatures. Turbulent velocities of 1.5 \kms\ and infall velocities of $-1.5$ \kms\ provided the best fit to the lines. The addition of a low speed molecular outflow (velocity a few times the infall speed) is included for completeness but is in fact not constrained well by the observational data: the Mopra beam is comparable to the size of the entire core and the outflow material will thus only occupy a small fraction of the beam, as evident from the presence of outflow line wings in the ATCA \ce{HCO+} data (Figure \ref{fig:atca_HCOp_blue}).

The chemical abundances used were those derived from the analysis of the observations in \citet{Lo2007}. For all molecules depletion, due to the freeze-out the molecules onto dust grains, is required in order to fit the data. However, the degree of depletion required varies widely between the species. \ce{^{13}CO} and \ce{C^{18}O} require a reduction of an order of magnitude of their gas phase abundance within $r < 0.9R$, from $1 \times 10^{-5}$ and $8 \times 10^{-7}$ to $1 \times 10^{-6}$ and $8 \times 10^{-8}$ for \ce{^{13}CO} and \ce{C^{18}O} respectively (Table \ref{tab:model_parameters}). This is entirely consistent with the presence of \ce{N2H+} in the centre of the core which should be destroyed by CO and is also consistent with integrated intensity maps of these species which show the CO isotopes are only weakly present above the background molecular cloud. More modest depletion is implied for the \ce{HCO+} data and the modelling suggests that this species is depleted within $r < 0.3R$. Little freeze-out of CS is implied by the modelling which calls for a depletion only within the central regions at $r < 0.1R$. The core (envelope) and outflow parameters of the best fit model are summarised in Table \ref{tab:model_parameters}, the modelled spectra are shown in Figure \ref{fig:cc_mopra_model}.

\subsection{Modelling of the ATCA data} \label{sec:modelling_atca}
\begin{figure}
  \centering
  \includegraphics{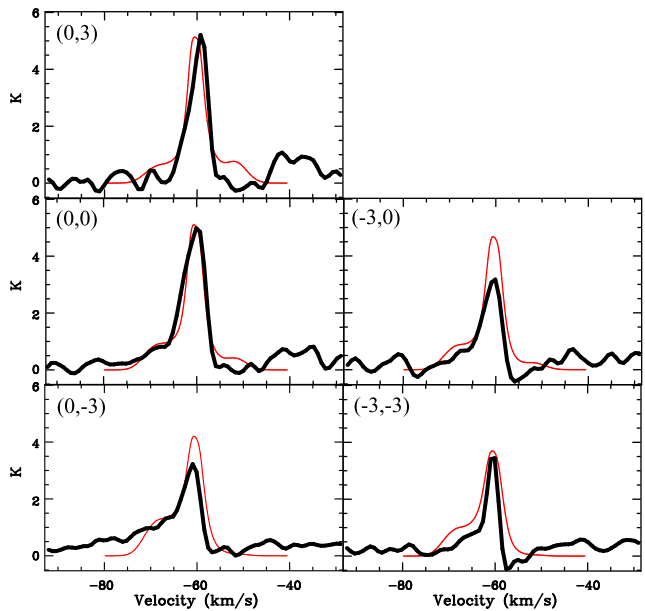}
  \caption{Modelled \ce{HCO+} spectra (thin lines) overlaid on the observed spectra (ATCA; thick lines) across the core. The numbers in top left brackets are offsets from the \ce{HCO+} peak in arcseconds.}
  \label{fig:HCOp_model_atca}
\end{figure}

Based on the best fit model of Mopra data, we further refined the input parameters to replicate the blue-shifted line wings found in the ATCA \ce{HCO+} line profiles. We adopted an `hourglass'-shaped outflow model \citep{Rawlings2004} with an opening angle of 50 degrees. In order to reproduce the centre line intensity, the core density has to be doubled from the Mopra data model (Table \ref{tab:model_parameters}). The outflow density is also an order of magnitude higher and temperature of 50 K to give the line wings intensity, this is consistent with the suggestion that the Mopra \ce{HCO+} data suffers from beam dilution as the source does not fill up the beam, and that the interferometric \ce{HCO+} data has filtered out the extended structure of the core. The extent of the \ce{HCO+} line wings are reproduced with an outflow velocity of 12 \kms, while a turbulent velocity of 3.0 \kms\ gives the smooth blending profile between the centre line component and the outflow line wings; the outflow is oriented such that the blue-shifted outflow lobe is south-west of the source and the red-shifted lobe falls outside of the primary beam. The \ce{HCO+} abundance has to be increased by two orders of magnitude in order to reproduce the observed intensity. This is consistent with the findings of \citet{Rawlings2004} that \ce{HCO+} abundance is enhanced in outflows. Shown in Figure \ref{fig:HCOp_model_atca} are sample spectra of the observed ATCA \ce{HCO+} overlain on the model spectra.

\begin{table}
\caption{Summary of characteristic model parameters for the ATCA and Mopra data determined using the {\sc MOLLIE} radiative transfer code. All parameters vary with functional forms that are described in detail in the text. Density values listed here are the central densities following a density distribution of a simple Plummer sphere as detailed in Section \ref{sec:modelling}. The listed values are uncertain by a factor of two, as derived from the number of plausible models.}
\label{tab:model_parameters}
\begin{minipage}{0.5\textwidth}
\centering
\begin{tabular}{lll}
  \hline
	Mopra \ce{HCO+}, \ce{CS}, \ce{C^{18}O} and \ce{^{13}CO} & Envelope & Outflow \\
   & + Core\footnote{Mopra beam resolution covers both the core and envelope.} & \\
	\hline
  \hline
  Density ($\times 10^5 \, \rm{cm^{-3}}$) & 5.0 & 0.1 \\
  Velocity (\kms) & $-1.5$\footnote{Infall velocity.} & 3.0\\
  Turbulent width (\kms) & 1.5 & 1.5\\
  Temperature (K) & 17 & 17 \\
  Abundance \ce{C^{18}O}/\ce{H2} ($\times 10^{-7}$) & 8.0 & 8.0 \\
  Abundance \ce{^{13}CO}/\ce{H2} ($\times 10^{-5}$) & 1.0 & 1.0 \\
  Abundance \ce{CS}/\ce{H2} ($\times 10^{-10}$) & 4.0 & 4.0 \\
  Abundance \ce{HCO+}/\ce{H2} ($\times 10^{-10}$) & 6.8 & 6.8 \\
	\hline
	ATCA \ce{HCO+} & Core & Outflow\\
 	\hline
  \hline
  Density ($\times 10^5 \, \rm{cm^{-3}}$) & 10.0 & 3.0 \\
  Velocity (\kms) & $-1.5^b$ & 12.0\\
  Turbulent width (\kms) & 3.0 & 3.0\\
  Temperature (K) & 17 & 50 \\
  Abundance \ce{HCO+}/\ce{H2} ($\times 10^{-10}$) & 6.8 & 34.0 \\
  \hline
\end{tabular}
\end{minipage}
\end{table}

In summary, a self-consistent radiative transfer model was generated using the size, temperature, density and abundances reported in \citet{Lo2007}. Supersonic infall and turbulence and heavy depletion of CO, due to freeze-out, is required to produce a good fit to the line widths, shapes and strengths.

\section{Discussion} \label{sec:discussion}

\core\ is located in the G333 giant molecular cloud complex which has been the subject of a multi-molecular line survey project carried out by the authors. Several papers of results have now been published from this work \citep{Bains2006, Wong2008, Lo2009} and so it is now possible to compare \core\ with other sites of high mass star formation within G333.

Models of high mass star formation emphasise the role of turbulence \citep{McKee2007}, so quantifying the variation in the properties of turbulence throughout G333 is important to the testing of such models. Turbulent velocities in \core\ are highly supersonic and large by low mass star formation standards. However, they are significantly less than those measured in the three most massive star forming regions in G333 ($\sim 6$ \kms\ ; Bains et al, in prep.). The core has a lower mass and a lower turbulent (and infall) velocity that might be expected simply from virial theorem arguments, where the mass of the cold core is $2.2 \times 10^3$ \solarM\ \citep{Garay2004}, an order of magnitude lower than the three most massive sources in G333. 

The coincidence of the continuum sources at four different wavelengths (3 and 7 mm, 24 and 70 \um) further confirms that the internal heating source is deeply embedded in the cold core. Whether the continuum emission is due to warm dust or free-free emission from a deeply embedded H{\sc ii} region can be examined using the spectral index $\alpha$ (see Section \ref{sec:results}). The three millimetre wavelength (1.2, 3 and 7 mm) data points yield a spectral index $\alpha$ of $4.2 \pm 0.07$ and dust emissivity $\beta$ of 2.2, which is comparable to the dust emissivity of $\beta = 2$ found by \citet{Hill2006} for cold cores. Thus, to at least 7-mm wavelength, the continuum emission is due to thermal dust emission rather than free-free emission. At longer wavelengths, there is no detectable (free-free) continuum emission from the core, thus it is possible that an H{\sc ii} region still has not yet developed at this stage. 

The secondary molecular source (denoted `B' in Figure \ref{fig:intensity}) has no detectable millimetre, 24 nor 70 \um\ continuum emission, thus no indication of an internal heating source. We speculate this source could be an even less evolved object.

Contrary to the Mopra \ce{HCO+} spectra profiles, the high resolution ATCA data does not show infall asymmetry. This is likely due to extended emission, such as the contribution from the envelope, where infall takes place, is filtered out in the interferometry data. 

Two possible explanations for the absence of outflow line wings in Mopra \ce{HCO+} line profiles are lack of sensitivity and/or the outflow (spatial) scale. The ATCA \ce{HCO+} data has a spatial resolution of $6 \times 3$ square arcseconds while the Mopra data has resolution of 36 arcseconds, if the outflow is not widely spread and only partially fills the Mopra beam, then it is heavily diluted.  Thus line wings are not seen in the Mopra spectra profiles. An abundance enhancement is also supported by the modelling, which requires a factor of 5 larger \ce{HCO+} abundance for the outflow component to reproduce well the higher resolution ATCA data than it does for the Mopra data.

\section{Conclusions} \label{sec:conclusions}
\core\ is an excellent example of the very early stages of massive star formation. The cloud is cold, moderately turbulent and possesses an outflow which indicates the presence of a central driving source. \ce{HCO+} is enhanced in the outflow. The spectral index of the primary source indicates the continuum emission at wavelengths shorter than 7-mm is due to thermal dust emission. Undetectable continuum emissions at longer wavelengths suggests the absence of free-free emission from ionised gas, thus an H{\sc ii} region has not yet developed. We have modelled the molecular emission using the 3D radiative transfer code {\sc MOLLIE}, obtaining the best match to the data with a model where there is a massive, dense core containing an embedded source driving an outflow, but without the presence of detectable ionized gas from an H{\sc ii} region.

\section*{Acknowledgments}
NL acknowledges partial support from Center of Excellence in Astrophysics and Associated Technologies (PFB 06) and Centro de Astrof\'{i}sica FONDAP\,15010003. NL's postdoctoral position at CEA/Irfu was funded by the Ile-de-France Region. MPR acknowledges support from a Science Foundation Ireland Research Frontiers grant. PAJ acknowledges partial support from Centro de Astrof\'{i}sica FONDAP\,15010003 and the GEMINI-CONICYT FUND. This work was supported by a Science Foundation Ireland Research Frontiers grant. The Mopra Telescope and ATCA are part of the Australia Telescope and are funded by the Commonwealth of Australia for operation as National Facility managed by CSIRO. The UNSW-MOPS  Digital Filter Bank used for the observations with the Mopra Telescope was provided with support from the University of New South Wales, Monash University, University of Sydney and Australian Research Council. The NANTEN2 Observatory is a collaboration between Nagoya University, Osaka University, KOSMA, Universit\"at zu K\"oln, Argelander-Institet Universit\"at Bonn, Seoul National University, ETH Z\"urich, University of New South Wales and Universidad de Chile. This research has made use of the NASA/IPAC Infrared Science Archive which is operated by the Jet Propulsion Laboratory, California Institute of Technology, under contract with NASA.

\bibliographystyle{mn2e}
\bibliography{coldcore}


\bsp

\label{lastpage}

\end{document}